\documentclass[final,3p,times,twocolumn]{elsarticle}

\usepackage{amssymb}
\usepackage{amsmath}
\usepackage{algorithm}
\usepackage{algpseudocode}
\usepackage{graphicx}
\usepackage{url} 
\usepackage{siunitx} 
\usepackage{hyperref}

\journal{}

\begin{document}

\begin{frontmatter}



\title{A Novel Scalable High Performance diffusion solver for multiscale cell simulations}

\author[BSC]{Jose-Luis Estragues-Muñoz} \ead{jose.estragues@bsc.es}
\author[BSC,UPC]{Carlos Alvarez} \ead{carlos.alvarez@bsc.es}
\author[BSC,CSIC]{Arnau Montagud} \ead{arnau.montagud@csic.es}
\author[BSC,UPC]{Daniel Jimenez-Gonzalez} \ead{daniel.jimenez@bsc.es}
\author[BSC,ICREA]{Alfonso Valencia} \ead{alfonso.valencia@bsc.es}
\cortext[cor1]{Corresponding author}

\affiliation[BSC]{organization={Barcelona Supercomputing Center (BSC)}
}
\affiliation[UPC]{organization={Universitat Politecnica de Catalunya (UPC)},
            addressline={Carrer de Jordi Girona Edifici B6},
            city={Barcelona},
            postcode={08034},
            country={Spain}
}

\affiliation[CSIC]{organization={Institute for Integrative Systems Biology (I2SysBio), CSIC-UV}
}

\affiliation[ICREA]{organization={ICREA},
            addressline={Passeig de Lluís Companys, 23},
            city={Barcelona},
            postcode={08010},
            country={Spain}
}

\begin{abstract}
Agent-based cellular models simulate tissue evolution by capturing the behavior of individual cells, their interactions with neighboring cells, and their responses to the surrounding microenvironment. 
An important challenge in the field is scaling cellular resolution models to real-scale tumor simulations, which is critical for the development of digital twin models of diseases and requires the use of High-Performance Computing (HPC) since every time step involves trillions of operations. 
We hereby present a scalable HPC solution for the molecular diffusion modeling using an efficient implementation of state-of-the-art Finite Volume Method (FVM) frameworks. The paper systematically evaluates a novel scalable Biological Finite Volume Method (BioFVM) library and presents an extensive performance analysis of the available solutions. Results shows that our HPC proposal reach almost 200x speedup  and up to 36\% reduction in memory usage over the current state-of-the-art solutions, paving the way to efficiently compute the next generation of biological problems.
\end{abstract}



\end{frontmatter}


%

\section{Introduction}

Agent-based multiscale simulators are essential tools for modeling complex biological systems, providing valuable insights into cellular behavior and their interactions with the surrounding microenvironment. One of the most promising goals in this field is the creation of Human Digital Twins (HDT), virtual patient models that simulate disease progression and support personalized treatment strategies. However, these simulations involve heavy computational workloads, especially as they scale from tissue-level models with approximately 10$^{6}$ cells to organ-level models involving 10$^{12}$ cells \cite{montagud2021systems}
and need optimised simulation codes.\\ 

HDT simulations require models that capture biological systems at the centimeter scale —corresponding to the size of tumors— while maintaining micrometer resolution to accurately simulate cellular processes. Multiscale cell simulations tackle this problem by addressing diverse biological processes at different time scales\cite{montagud2021systems}. Among these processes, the diffusion-decay of substrates (such as nutrients and drugs) is the most computationally demanding as it is simulated most frequently and needs to solve a large number of differential equations at each time step.\\

PhysiCell \cite{Ghaffarizadeh2018}, a physics-based multiscale cell simulator, operates at the micrometer resolution and uses the BioFVM \cite{Ghaffarizadeh2016} library to model the diffusion-decay of substrates in the microenvironment. Both PhysiCell and BioFVM are open-source, written in C++ and parallelized using OpenMP. However, as researchers work to scale these simulations to centimeter-sized models, new challenges emerge. For example, the MPI-based extension BioFVM-X \cite{Saxena2021} was introduced to enable simulations of real-sized tumors and potentially entire organs, but revealed inefficiencies in both execution time and scalability.\\

One of the primary computational bottlenecks in these large-scale simulations lies in the diffusion-decay step, as it employs the Tridiagonal Matrix Algorithm (TDMA) to solve large sets of partial differential equations (PDEs) across the three spatial dimensions \cite{Ghaffarizadeh2016}. The current implementations suffer from memory limitations in shared memory systems and serialization issues between MPI processes in distributed systems, which hinders scalability, as observed in performance results from BioFVM-X \cite{Saxena2021}. Furthermore, as simulation frameworks evolve to include more complex models of cell behavior —such as PhysiBoSS \cite{letort2019physiboss, PonceDeLeon2023}, which uses stochastic Boolean networks to simulate cell phenotypes, and PhysiMess \cite{Noel2023PhysiMeSS}, which models the extracellular matrix, the demand for a more efficient computational infrastructure becomes even more critical.\\

In this paper, we present BioFVM-B, an open-source high-performance computing solution designed to overcome the limitations of BioFVM and BioFVM-X in modeling microenvironments for multiscale cell simulations. BioFVM-B is a hybrid MPI-OpenMP parallel SIMD application that follows a data decomposition strategy that significantly enhances the efficiency of the diffusion-decay process. BioFVM-B improves execution times while maintaining accuracy, achieving accelerations of up to 34.8× over BioFVM and 196.8× over BioFVM-X. Results show that BioFVM-B overcomes state-of-the-art diffusion solvers both in performance and scalability on Marenostrum 5 \cite{MareNostrum5}. BioFVM-X is capable of simulate a full organ microenvironment, underscoring its potential to enable cellular resolution of Human Digital Twins (HDT) for personalized medicine. \\

The article describes the algorithm and complexity of the diffusion-decay time step in Section \ref{cap:intro_solver}. Then, Section \ref{cap:novel_solver} presents our proposal to exploit the parallelism at different levels: data level parallelism with a novel vector implementation, thread level parallelism using OpenMP and distributed parallelism using MPI. Experimental setup and evaluation results are described in Section \ref{cap:experimental} and \ref{cap:results} respectively.

\section{Related work}\label{cap:intro_solver}

The physics-based cell simulator PhysiCell utilizes the BioFVM library to model substrate diffusion and decay by solving a set of partial differential equations (PDEs) at each time step. The diffusion-decay process in 3D is modeled using an additional decomposition along the \textit{x}-, \textit{y}-, and \textit{z}-dimensions. This decomposition is facilitated by the Locally One-Dimensional (LOD) method, which simplifies computational complexity while preserving accuracy. The solver has demonstrated first-order accuracy in time, producing reliable results for time steps ranging from 0.01 to 0.1 minutes \cite{Ghaffarizadeh2018}.\\ 

The solver operates in 7 stages: four applications of Dirichlet boundary conditions, followed by the solution of the diffusion-decay process along each dimension. Algorithm \ref{alg:x_diff} exemplifies for the \(x-\)dimension. Each spatial dimension is decomposed into multiple independent tridiagonal systems of equations, with system size determined by the other two spatial dimensions and the number of substrates. \\

\begin{algorithm}
\caption{Diffusion-decay 3D solver 1 dimensional computation}
\label{alg:x_diff}
\begin{algorithmic}[1]
\For{$\text{y} \gets \text{y\_start} \ \textbf{to} \ \text{y\_end} \ \textbf{step} \ \text{dy}$}
    \For{$\text{z} \gets \text{z\_start} \ \textbf{to} \ \text{z\_end} \ \textbf{step} \ \text{dz}$}
        \State{TDMA\_forward\_sweep(y,z)}
        \State{TDMA\_backward\_sweep(y,z)}
    \EndFor
\EndFor
\end{algorithmic}
\end{algorithm}

The Tridiagonal Matrix Algorithm (TDMA), also known as the Thomas algorithm, is used to efficiently solve each system of equations using a simplified form of Gaussian elimination \cite{thomassolver}. The TDMA is an algorithm with reduced memory overhead using the sparsity of the matrix, precomputed constant coefficients (as described in \ref{cap:a_coefficients}), and efficient serial resolution of cost O(n) with \textit{n} equations that compose each system. \\

Nevertheless, one potential drawback of this algorithm is its inherently serial nature, which primarily affects the solution of individual systems. Much of the existing literature to optimise TDMA focuses on developing parallelized implementations for solving a single tridiagonal system in parallel \cite{xian-he, kkim}. However, these approaches are not well-suited for our case, where the computational workload rises from solving multiple independent tridiagonal systems.\\

The independence of the tridiagonal systems allows us to leverage parallelization at a higher level, where each system can be assigned to separate computing units and techniques such as vectorization, can be applied. This approach ensures optimal performance in our application while maintaining accuracy.\\

\subsection{Algorithm complexity and number of operations}

The number of systems of equations solved at each diffusion-decay time step is determined by the product of the number of substrates and  the number of voxels, the unit of discretization in the Finite Volume Method (FVM), in the other two dimensions. Equation \ref{eq:numTD} shows the number of tridiagonal systems to be solved per diffusion iteration, where \textit{S} represents the number of substrates, \textit{X,Y,Z} the size of the bounding box and $\Delta$\textit{X},$\Delta$\textit{Y},$\Delta$\textit{Z}, the discretization value for each of the dimensions. \\

\begin{equation} 
\label{eq:numTD}
     TDeqs = S * (\overbrace{\frac{Y}{\Delta Y}\frac{Z}{\Delta Z}}^{X-diffusion} + \overbrace{\frac{X}{\Delta X}\frac{Z}{\Delta Z}}^{Y-diffusion} + \overbrace{\frac{X}{\Delta X}\frac{Y}{\Delta Y}}^{Z-diffusion})
\end{equation}

TDMA is composed of two sequential stages with different numbers of operations. During \textit{Forward} sweeping, the first iteration simplifies the first equation by dividing by \textit{b$_{1}$}, while the remaining iterations perform an \textit{axpy} and a division. During \textit{Backward} sweeping, a negative \textit{axpy} operation is performed in all iterations except for the last one. \\

The number of arithmetic operations to solve the TDMA can be computed by:

\begin{equation} \label{eq:eqop}
    \resizebox{\columnwidth}{!}{$
     Ops(voxels) = \overbrace{voxels + axpy*(voxels-1)}^{Forward} + \overbrace{naxpy*(voxels - 1)}^{Backward}   
    $}
\end{equation}

Therefore, the total number of floating point operations combines the total number of tridiagonal systems (equation \ref{eq:numTD}) and the number of operations per system (equation \ref{eq:eqop}) in equation \ref{eq:flop}.

\begin{equation} \label{eq:flop}
    \resizebox{\columnwidth}{!}{$
      Flop = substrates * (\overbrace{Ops(\frac{X}{\Delta X})*\frac{Y}{\Delta Y}\frac{Z}{\Delta Z}}^{X-diffusion} + \overbrace{Ops(\frac{Y}{\Delta Y})*\frac{X}{\Delta X}\frac{Z}{\Delta Z}}^{Y-diffusion} + \overbrace{Ops(\frac{Z}{\Delta Z})*\frac{X}{\Delta X}\frac{Y}{\Delta Y}}^{Z-diffusion})    
    $}
\end{equation}

The algorithm complexity can be deduced from equation \ref{eq:flop}. Assuming a microenvironment with equal sides (X=Y=Z) and same discretization value ($\Delta$X = $\Delta$Y = $\Delta$Z), then the algorithm complexity is defined by the number of substrates (S) and the number of voxels per side (N = $\frac{X}{\Delta X}$ = $\frac{Y}{\Delta Y}$ = $\frac{Z}{\Delta Z}$). The TDMA has complexity $\mathcal{O}(N)$. Our algorithm complexity is then defined by equation \ref{eq:complexity}.
\begin{equation} \label{eq:complexity}
\begin{aligned}
    \mathcal{O}( S * (3N^{3})) \rightarrow \mathcal{O}( S N^{3})\hspace{1cm}\\
\end{aligned}
\end{equation}

Agent Based Modeling (ABM) requires efficient leverage of advanced computing technologies and biomedical research\cite{SEGMENT2015}. This is because this modeling approach has only recently started to address efficiency issues by tackling larger, more complex simulations \cite{montagud2021systems}. \\

There are different approaches to defining agent-based models (ABMs) for cellular simulations (more comprehensive reviews can be found elsewhere \cite{Metzcar_2019_review}). All of these tools simulate the diffusion of freely-roaming substrates by solving the equation corresponding to Fick's second law of diffusion using different methodologies that have been parallelized.\\

Some tools use the Finite Difference Method (FDM) to solve the diffusion equation. For instance, BioDynaMo \cite{biodynamo} incorporates its own solvers and optimizations targeting agent-based workloads in single-node executions. Timothy \cite{timothy} utilizes the Hypre library for solving FDM systems. SEGMENT\_HPC \cite{SEGMENT2015}, based on the Repast modeling tool \cite{Repast_2013}, uses a parallelized and distributed FDM to simulate large-scale systems. The main reason for not considering FDM in this work is that, unlike the Finite Volume Method, it does not inherently conserve mass, which is essential for accurately modeling biologically realistic microenvironments.\\

Some other tools use the Finite Element Methods (FEM) to solve the diffusion equation. CellSys \cite{tisim} applies an Euler forward method, which requires a very small diffusion time step, thus limiting its efficiency. Chaste \cite{chaste} uses FEM to solve diffusion equations and combines it with the flexibility of using different ABM types (off-lattice, on-lattice, Cellular Potts, etc). However, it requires solving large sparse systems and assuring mass conservation increases complexity.\\

PhysiCell\cite{Ghaffarizadeh2018} uses its own library, BioFVM\cite{Ghaffarizadeh2016} to parallelize the FVM in user-defined spatial resolutions. In addition, BioFVM-X\cite{Saxena2021} distributes the computation by introducing a one-dimensional domain partitioning scheme among different computing nodes and allows larger simulations. However, its  performance does not scale efficiently when increasing the number of compute nodes.\\

To enhance the computational performance of this task, variants of the TDMA have been designed to exploit parallelism in solving a single system of equations across diverse computing architectures. Previous studies have evaluated well-known tridiagonal solvers such as Cyclic Reduction (CR), Parallel Cyclic Reduction (PCR), and Recursive Doubling (RD) on GPUs\cite{gpusolvers}. FPGA implementations have also leveraged hardware parallelism through batched Thomas-based solvers \cite{fpgasovlers}. Furthermore, a divide-and-conquer strategy is employed in \cite{kkim} to enable parallelization for solving large-scale, distributed tridiagonal systems. While these approaches focus on solving a single system, our work targets the parallel solution of multiple such systems.\\

Different works have already studied the distribution of FVM to HPC in other fields, such as computer fluid dynamics\cite{Rio2021, Delis2009} or smoothed particle hydrodynamics\cite{Oger2016}. Nevertheless, none of this works is directly transferable to the multiscale models.

\section{BioFVM-B} \label{cap:novel_solver}
Our hybrid MPI-OpenMP implementation introduces novel optimizations for existing libraries BioFVM and BioFVM-X and is designed to address their inefficiencies by providing an HPC-oriented implementation. These improvements include the adoption of a more appropriate data structure for storing microenvironment substrates' values and refined diffusion-decay solvers that better exploit the inherent parallelism of the algorithm.\\

\subsection{Distributed Microenvironment class}
\begin{figure}[h]
    \centering
    \includegraphics[width=\linewidth]{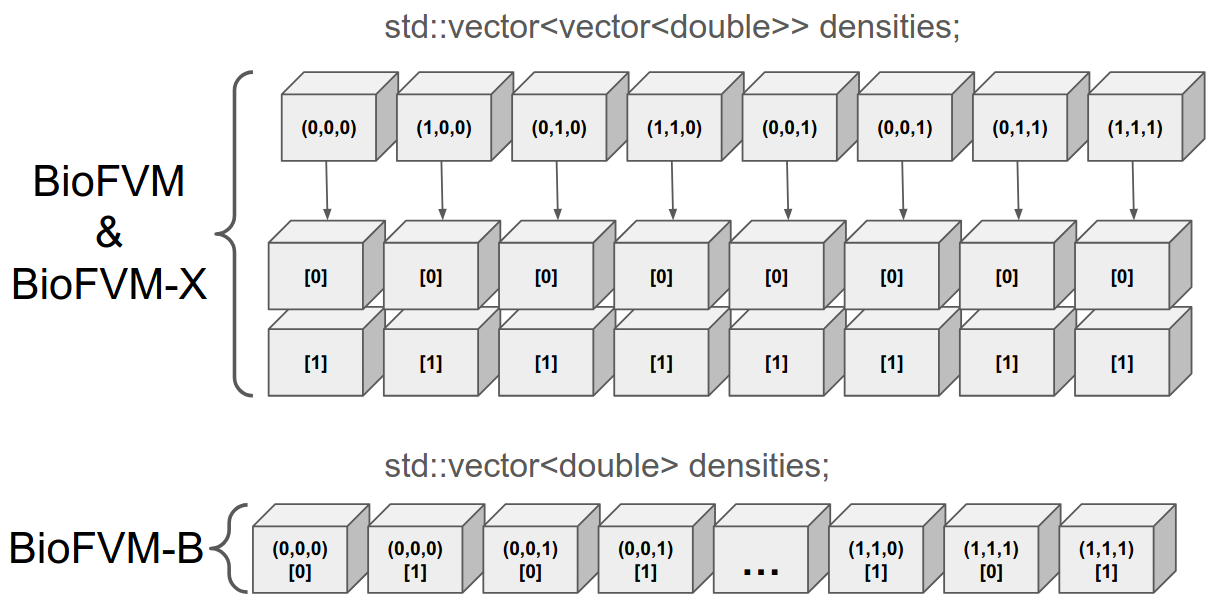}
        
    \caption{Comparison of the data structures used by BioFVM and BioFVM-X (top panel) vs. BioFVM-B (bottom panel) for storing substrate values of the microenvironment, under the variable name \textit{p\_density\_vectors} within the \textit{Microenvironment} class. The figure illustrates an example of a cubic microenvironment with dimensions of 2x2x2 voxels, modeling 2 substrates. Positions within the microenvironment are denoted by \((x, y, z)\), while substrate values are indicated as \([id]\).}

    \label{fig:data_structure}
\end{figure}

The design of data structures is crucial for efficiently managing increasingly complex microenvironment simulations. In existing implementations such as BioFVM and BioFVM-X, microenvironment density values are organized within a data structure composed of a vector of vectors. Here, the first vector encodes the spatial coordinates layout of \(x\), \(y\), and \(z\) axes (Figure \ref{fig:data_structure}, top panel). For each element of this vector, a referenced vector stores the density values of various substrates corresponding to the specific spatial position. Accessing each referenced vector results in slower value retrieval and memory usage. 

Our proposal simplifies this architecture by utilizing a contiguous memory allocation of the four-dimensional vector which improves the representation of microenvironment data. The structure of this vector integrates density values with the \(z\), \(y\), and \(x\) coordinates, (Figure \ref{fig:data_structure}, lower panel). This strategic modification is designed to enhance memory access patterns during the execution of the diffusion-decay solver, thereby optimizing computational efficiency. Furthermore, this approach aligns the data distribution more closely with the domain partition of both BioFVM-X and BioFVM-B. Another critical advantage of this implementation is the reduction of indirections and the elimination of extra headers associated with the double vector structure, resulting in relevant memory savings. \\

\subsection{Diffusion-decay solver for distributed computing} \label{se:blocks}
\begin{figure}[h]
    \centering
    \includegraphics[width=\linewidth]{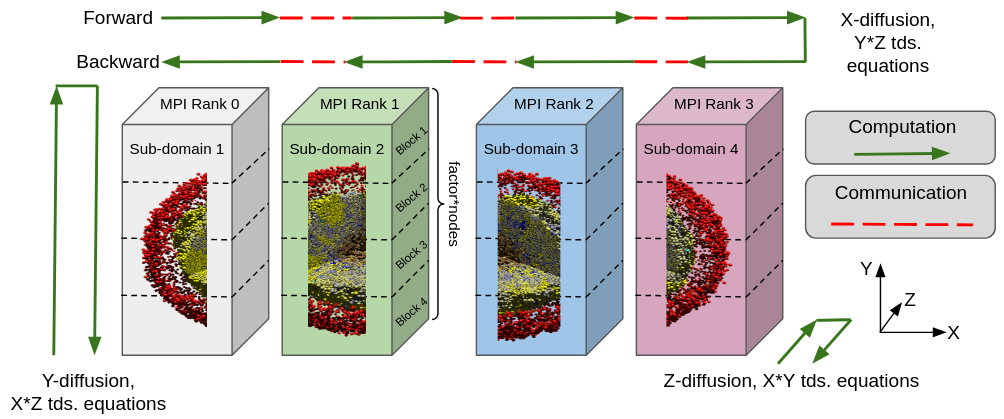}
    \caption{BioFVM-B diffusion-decay solver data distribution. Access patterns are shown for each of the dimension computations. Example of the division of the simulation domain into 4 subdomains that are mapped to different MPI processes. The blocking feature is introduced by BioFVM-B. Illustrated example of 4 MPI processes and \textit{factor} 1.}
    \label{fig:data_decomposition}
\end{figure}

A novel diffusion-decay solver is introduced in BioFVM-B to enhance computational performance and scalability enabling larger, more complex, simulations. The solver traverses each of the 3 dimensions using the TDMA as described in section \ref{cap:intro_solver}. Distributing the microenvironment into different processes requires adapting the solver to include communication using MPI. In the case of BioFVM-B, inherited from BioFVM-X, this consists of a 1-dimensional partition of the simulation domain (Figure \ref{fig:data_decomposition}). Each dimension's diffusion computation has a distinctive access data pattern which consists of traversing the same dimension from the initial coordinate to its final coordinate, in a forward phase, and reverse, backward phase. Subdomain communication is required when solving the distributed \(x\)-dimension. The right \(x\)-coordinate plane is sent to the next MPI process during \textit{Forward} elimination and the left \(x\)-coordinate  is sent to the previous MPI process for the \textit{Backward} elimination.\\

In order to exploit the parallelism, we process x-diffusion by blocks, to allow overlapping communication and computation. Blocks are composed of contiguous coordinates from the starting \(x\)-coordinate plane and are processed sequentially and sent to the next process. When the solver starts, the process with an ID equal to 0 is responsible for processing blocks and sending the frontier \(x\)-plane to the following process while the rest of the processes wait. Concurrent computation and communication of the blocks is included by using \textit{non-blocking} MPI directives. In addition, the BioFVM-B microenvironment data structure enables the direct mapping of the blocks to MPI buffers, reducing the cost of parsing data from MPI buffers to the layout of the density values and vice versa. \\

 Adjusting the number of blocks (NB) used has direct effects on the solver's execution time, as it affects the number of messages between subdomains and the processing time per block (see section \ref{cap:results}). Additionally, the workload to be divided into blocks is determined by the simulation domain size which is only known at the simulation setup. Thus, we propose a heuristic capable of handling all possible simulation scenarios in equation \ref{eq:blocks}. This approach proposes a number of blocks based on the MPI processes and a constant \(k\), called \textit{factor}, that can be adjusted depending on the problem size.\\
\begin{equation}
\label{eq:blocks}
    Number \: of \: blocks(nodes) = k * nodes
\end{equation}

\subsection{Vectorized solvers} 
BioFVM-B has been adapted to exploit data level parallelism using Single Instruction Multiple Data (SIMD) instructions efficiently. Considering the novel microenvironment data structure, the execution sequence has been reordered to increase the number of instructions that access operands stored in consecutive memory addresses, which varies across dimensions and results in different vector length potentials. This modification exploits the concurrent resolution of the multiple systems.\\

X-diffusion offers the largest vector length potential, making it ideal for accelerating block computations. The \textit{for loop} at line 2 of Algorithm \ref{alg:x_vec} initializes the coordinates of all systems to be solved, and can be parallelized using a combination of OpenMP threads and SIMD instructions. This kernel is vector-agnostic, meaning the vector length cannot exceed the size of the space defined by the \textit{y} and \textit{z} dimensions and the number of substrates.\\

\begin{algorithm}
\caption{BioFVM-B vectorized solver x-diffusion}
\label{alg:x_vec}
\begin{algorithmic}[1]
\State{OMP for loop}
\For{$\text{coordinates in (y,z,s) } \textbf{step } \text{VL}$}
    \State{TDMA\_forward\_X\_sweep()}
    \State{TDMA\_backward\_X\_sweep()}
\EndFor
\end{algorithmic}
\end{algorithm}

Y-diffusion can perform vector operations across the zz and substrate spaces. Algorithm \ref{alg:y_vec} illustrates the implemented parallelism: OpenMP threads distribute the \textit{x}-coordinates (line 2), and each thread applies TDMA using SIMD instructions to concurrently solve the systems spanning the \textit{z} and substrate dimensions (line 3).

\begin{algorithm}
\caption{BioFVM-B vectorized solver y-diffusion}
\label{alg:y_vec}
\begin{algorithmic}[1]
\State{OMP for loop}
\For{$\text{coordinates in (x) }$}
    \For{$\text{coordinates in (z, s) } \textbf{step } \text{VL}$}
        \State{TDMA\_forward\_Y\_sweep()}
        \State{TDMA\_backward\_Y\_sweep()}
    \EndFor
\EndFor
\end{algorithmic}
\end{algorithm}

Assuming each dimension can contain a number of voxels from order 10$^{2}$ up to 10$^{5}$ for centimeters size and substrates up to 10$^{1}$, the number of operations per diffusion are: 10$^{4}$ up to 10$^{11}$ for X-diffusion, 10$^{2}$ up to 10$^{6}$ for Y-diffusion by merging operations along z-coordinates and different substrates, and only up to 10 for z-diffusion since it can be only applied to the number of substrates within the simulation. In contrast, BioFVM and BioFVM-X can only exploit vector operations on the order of substrates, around 10 operations, using the present assumptions. \\

\section{Experimental Setup} \label{cap:experimental}
\subsection{Hardware configuration}
All results have been obtained using the MareNostrum 5 supercomputer general-purpose partition. Each of the 6.192 nodes \cite{MareNostrum5} is equipped with 112 cores split across two sockets with one Intel Xeon Platinum 8480+  each, running at 2 GHz and 256 GB of main memory. The vectorized solver version was tested using SIMD AVX256D. Libraries were compiled with GCC 13.2.0 \cite{GCC13} and OpenMPI 4.1.5 \cite{OpenMPIv4}.\\

\subsection{Software configuration}
In the present work, we systematically compare three different, yet related, libraries that apply a Finite Volume Method to solve how substrates diffuse in a 3D environment. BioFVM 1.0.4\cite{Ghaffarizadeh2016} is an open-source shared memory implementation that uses OpenMP\cite{Ghaffarizadeh2016}, while BioFVM-X 1.0.1 is a distributed refactoring of BioFVM that includes MPI and OpenMP \cite{Saxena2021}. Lastly, BioFVM-B 1.0 in the present work inherits the domain partition from BioFVM-X and provides a scalable and optimised solution for simulations using MPI and OpenMP.\\

Additionally, we used performance analysis tools \cite{BSC_Tools} to study the execution behavior of these three libraries. Specifically, Extrae was used to obtain the execution traces \cite{Servat2013} and Paraver was used to analyze them \cite{Banchelli2020, Paraver_1995}.\\

\subsection{Benchmarks}
We have defined increasingly complex simulation configurations to study the performance of isolated functions of the code per time step. Simulations of cell populations vary greatly in size depending on the specific use case, ranging from a few micrometers for simulating cell properties to several centimeters. Since simulation size impacts the computational workload and the required resources, we propose six different configurations for performance testing that represent the range of magnitudes that the diffusion-decay solver usually handles in biomedical use cases (Table \ref{tab:configs}). The problem size is defined by the side length of the cubic microenvironment and the number of substrates involved in the simulation. For consistency and clarity, we fixed the spatial resolution of the FVM at $\Delta$x = $\Delta$y = $\Delta$z = 10 $\mu$m. The naming convention of the different problem configurations was selected with respect to the typical size of a human liver, which is roughly a 12 cm-side volume with hundreds of substrates \cite{liver2009}. This was used as a measure of how close is to simulate the diffusion of substrates in a full organ.

\begin{table}[h]
    \centering
    \begin{tabular}{|c|c|c|c|}
        \hline
        \multicolumn{1}{|c}{} & \multicolumn{2}{|c}{\textbf{Problem size}} & \multicolumn{1}{|c|}{} \\
        \hline
        \textbf{Name} & \textbf{Side ($\mu$m)} & \textbf{Substrates} & \textbf{GFlop} \\
        \hline
        4\% liver & 5000 & 2  & $\sim$4.7 \\
        \hline
        8\% liver & 10000 & 4 & $\sim$7.5  \\
        \hline
        12.5\% liver & 15000 & 8 & $\sim$41 \\
        \hline
        16.6\% liver & 20000 & 8 & $\sim$120 \\
        \hline
        21\% liver & 25000 & 8 & $\sim$1800 \\
        \hline
        100\% liver & 120000 & 1 & $\sim$256000 \\
        \hline
    \end{tabular}
    \vspace{0.2cm}
    \caption{Representative configurations parameters for BioFVM, BioFVM-X and BioFVM-B. The \textbf{GFlop} column counts the floating point operations needed to solve a single diffusion-decay time step.      
    }
    \label{tab:configs}
\end{table}

\section{Memory and Performance Analysis} \label{cap:results}

\subsection{Memory Requirement Limitations}
First, we wanted to test the capacity of the libraries for allocating microenvironments of increasing complexity. For this, we measured the Resident Set Size before and after the class \textit{Microenvironment} allocates the required problem size.\\

\begin{figure*}[h]
    \centering
    
    \includegraphics[width=\textwidth]
    {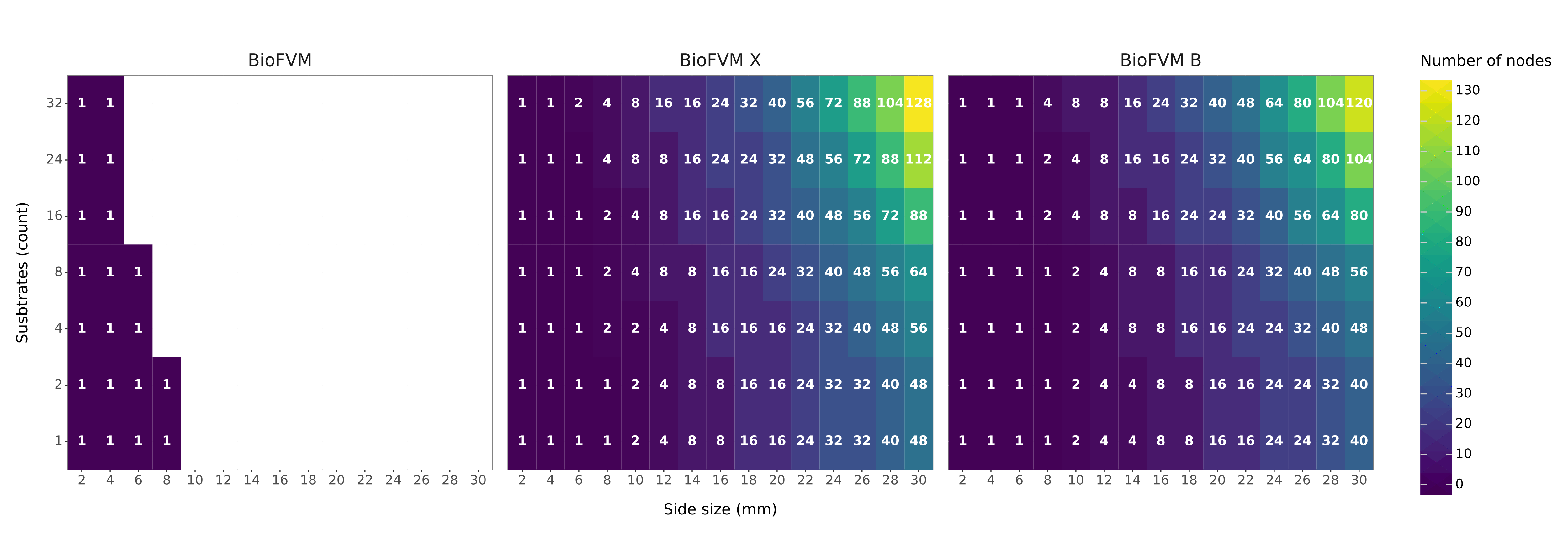}
        
    \caption{BioFVM, BioFVM-X and BioFVM-B MareNostrum 5 supercomputer's node usage to allocate different microenvironment sizes. }
    \label{fig:heatmap}
\end{figure*}

Figure \ref{fig:heatmap} compares the number of MareNostrum 5 computing nodes required by the three libraries to simulate various problem sizes. Due to its shared memory design, BioFVM has the most restrictions on the size of the domain and substrates it can model. BioFVM can only model domains of up to an 8-mm cube on a single machine, underscoring the limitations of shared memory architecture for scaling these simulations.\\

As expected, the distributed memory libraries BioFVM-X and BioFVM-B support simulations of much larger microenvironments, allowing for simulations at the centimeter-scale. For instance, simulating an 8-mm cube with 8 substrates requires 228 GB with BioFVM-B and 280.2 GB with BioFVM-X, while  BioFVM is not able to simulate this size using Marenostrum 5 nodes. \\

As microenvironment size increases, BioFVM-X requires more total memory than BioFVM-B. For example, simulating 1 substrate in a 3 cm microenvironment requires 48 nodes with BioFVM-X, whereas BioFVM-B requires 40 nodes. Columns 3 to 5 of table \ref{tab:results} summarizes memory usage across the selected problem sizes, emphasizing the considerable memory demands of these simulations and the reduced demand from BioFVM-B, showing a 36\% decrease in the 4\% liver case compared to BioFVM-X and 26.5\% decrease compared against BioFVM. \\

In summary, real-sized tumors of various centimeter-scale volumes demand memory capacity that is only provided by HPC clusters. The results of this section show that BioFVM-B solves the shared memory limitations of BioFVM and has a more efficient use of memory than BioFVM-X.

\begin{table*}[h]
    \centering
    \resizebox{\textwidth}{!}{
     \begin{tabular}{|c|c|
                    S[table-format=4.1] | S[table-format=4.2] | S[table-format=5.2]|
                    S[table-format=6.1] | S[table-format=6.0] | S[table-format=6.1]|}
        \hline
        \multicolumn{2}{|c}{} &  \multicolumn{3}{|c|}{\textbf{Resident Set Size [GB]}} & \multicolumn{3}{|c|}{\textbf{Mean execution time [ms]}}\\
        \hline
        \textbf{Name} & \textbf{Nodes} & \textbf{BioFVM} & \textbf{BioFVM-X} & \textbf{BioFVM-B} & \textbf{BioFVM} & \textbf{BioFVM-X} & \textbf{BioFVM-B} \\
        \hline
        4\% liver &  1 &  44  & 50.9  & 32.3   & 241.4 & 469 & 107.5\\
        \hline
        8\% liver &  8 &  \multicolumn{1}{S|}{\text{-}} & 453.4 &  304.7  & \multicolumn{1}{S|}{\text{-}} & 2390 & 270\\
        \hline
        12.5\% liver &  48 &  \multicolumn{1}{S|}{\text{-}} & 1833.3  & 1329.5   & \multicolumn{1}{S|}{\text{-}} & 21850 & 356.1\\
        \hline
        16.6\% liver &  96 &  \multicolumn{1}{S|}{\text{-}} & 4353.29  & 3179.54  & \multicolumn{1}{S|}{\text{-}} & 87050 & 453.4\\
        \hline
        21\% liver & 96 &  \multicolumn{1}{S|}{\text{-}} & 8507.85  & 6153.14 & \multicolumn{1}{S|}{\text{-}} & 163500 & 830.8\\
        \hline
        100\% liver & 125 &  \multicolumn{1}{S|}{\text{-}} & \multicolumn{1}{S|}{\text{-}}  & 27178.5 & \multicolumn{1}{S|}{\text{-}} & \multicolumn{1}{S|}{\text{-}} & 32943.2\\
        \hline
    \end{tabular}
    }
    \vspace{0.2cm}
    \caption{Evaluating the memory use and execution time of BioFVM, BioFVM-X, and BioFVM-B.    Results of the memory use and mean of 100 replicates of the diffusion-decay solver execution time. In these tests BioFVM-B used vectorisation with AVX256D instructions and a blocking \textit{factor} 2. The number of nodes was selected based on the minimum number of machines required for BioFVM-X to allocate the simulations.}
    \label{tab:results}
\end{table*}

\subsection{Performance results and scalability}
We performed a scalability analysis on the execution time of the diffusion-decay solver. First, we determined the optimal number of OpenMP threads for single-node executions and kept it fixed during the multi-node test runs. Second, we evaluated the performance of the solver in multi-node environments, focusing on the influence of the number of computational blocks on the execution time. Third, we compared the multi-node performance and scalability of BioFVM-X and BioFVM-B.

\subsubsection{Single node scalability}
Shared memory parallelism scalability was evaluated by increasing the number of OpenMP execution threads while maintaining the same workload. This analysis aims to identify the optimal number of cores that can be utilized within a single process, which can be up to 112 physical cores or 224 virtual threads by enabling Hyper-Threading on the MareNostrum 5 supercomputer.\\

\begin{figure}[h]
  \centering
  \includegraphics[width=\columnwidth]{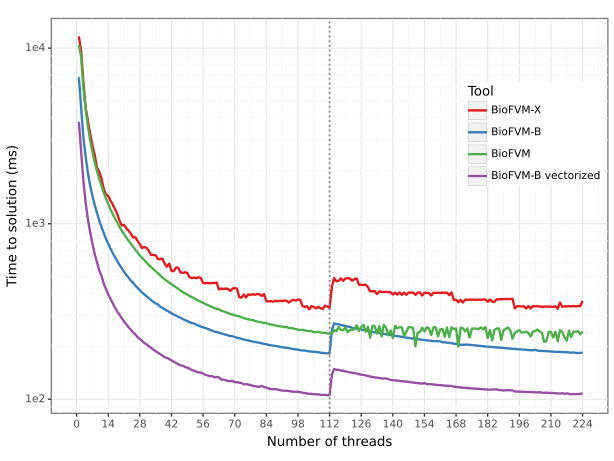}
  \caption{Comparison of time to solution per simulation time step for the diffusion-decay solver across BioFVM, BioFVM-X, and BioFVM-B implementations. Simulations are based on the 4\% liver test case. The vertical line indicates the number of physical cores; beyond this point, Hyper-Threading is utilized.} 
  \label{fig:t_ns}
\end{figure}

The vectorized solver in BioFVM-B is the fastest in single-node executions, as shown in Figure \ref{fig:t_ns}. In thread-to-thread comparisons, the vectorized BioFVM-B implementation outperforms all others, followed closely by the non-vectorized BioFVM-B version. Both clearly surpass the performance of the other implementations, underscoring the gains achieved through changing the data structure and adding vectorization. An AVX512D version of BioFVM-B was also evaluated; however, CPU frequency throttling limits its effectiveness, resulting in no execution time improvement over the AVX256D variant. In contrast, BioFVM-X demonstrates slower performance on a per-thread basis. 

\begin{figure}[h]
  \centering
  \includegraphics[width=\columnwidth]{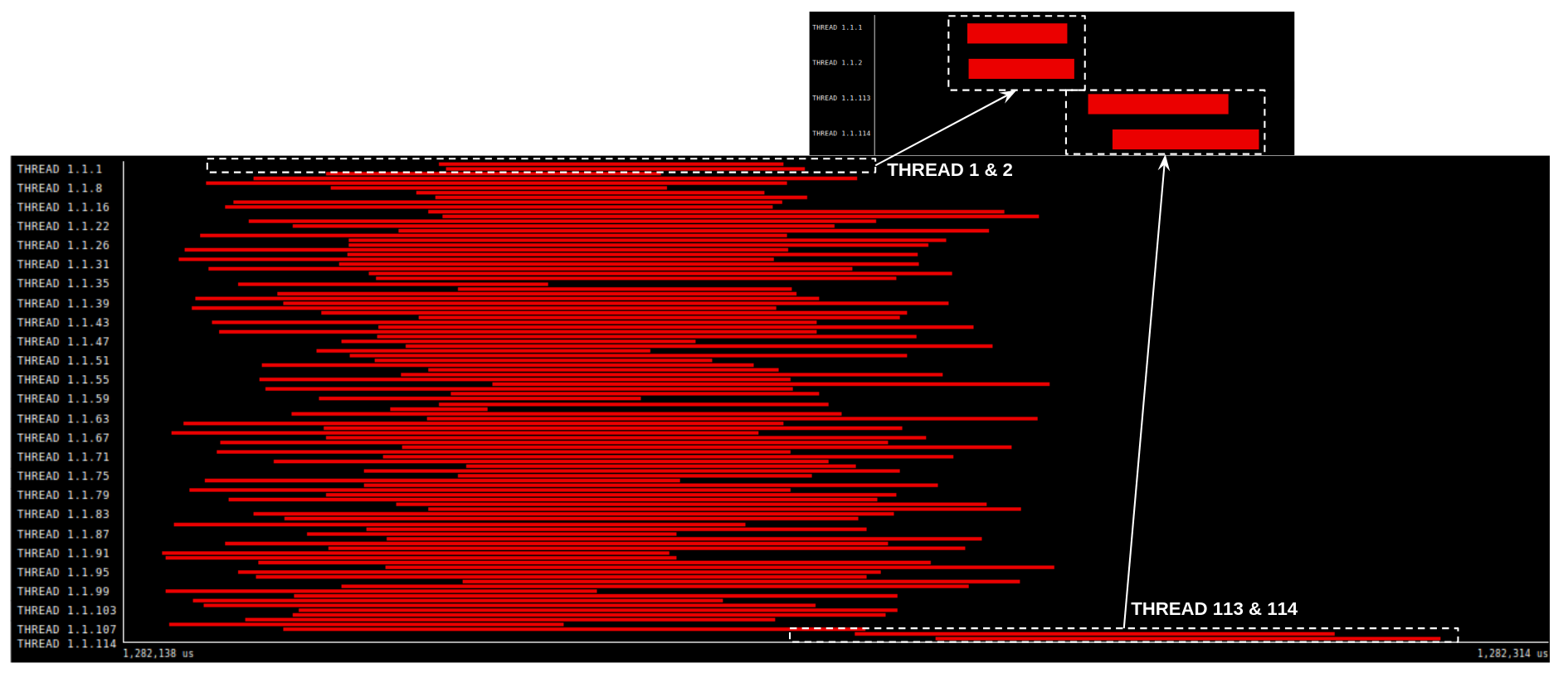}
  \caption{BioFVM-B diffusion-decay execution trace in MareNostrum 5 supercomputer analysed with BSC tools\cite{BSC_Tools}. Hyper-Threading serialization effects are shown in the threads sharing the same physical core.} 
  \label{fig:hyperthreading}
\end{figure}

Beyond 112 threads, enabling Hyper-Threading results in a decline in performance. Among the tested implementations, BioFVM is the only implementation that exhibits a modest performance improvement when all 224 threads are activated, increasing the speedup from 42.7× at 112 threads to 44.25×. In contrast, the MPI-based versions suffer performance degradation under Hyper-Threading due to execution serialization between thread pairs—specifically thread IDs {1,113} and {2,114}—that are mapped to the same physical core. The execution trace shown in Figure \ref{fig:hyperthreading} illustrates this phenomenon. The Y-axis represents threads, and the X-axis shows the progression of time. Red regions indicate periods of useful computation, whereas black regions denote idle thread states. Thread pairing causes contention for shared physical resources, ultimately reducing efficiency rather than improving it.\\

In summary, BioFVM-B has proven to be the most efficient library for single-node executions, demonstrating significant benefits from vectorization. The results indicate that the optimal configuration for performance is to use 112 physical cores, as Hyper-Threading degrades performance due to resource contention. Based on this, the following multi-node scalability tests were configured with 112 threads per node for all the tools.\\

\subsubsection{Blocking design space exploration}

The BioFVM-B diffusion-decay solver aims to address the serialization inefficiencies in BioFVM-X by grouping PDEs into independent groups, which are then processed serially (Section \ref{cap:novel_solver}). The number of blocks (NB) is a key parameter affecting the execution time of the xx-diffusion phase of the solver. However, the optimal number depends on several factors, such as the problem size, the number of computing nodes, and the cluster specifications (notably the available computing and communication resources). These factors influence the trade-off of increasing NB to enhance multi-node parallelism while simultaneously increasing the volume of MPI messages that need to be transmitted among nodes, which raises the overhead of processing the blocks.\\

\begin{figure}[h]
  \centering
  \includegraphics[width=\columnwidth]{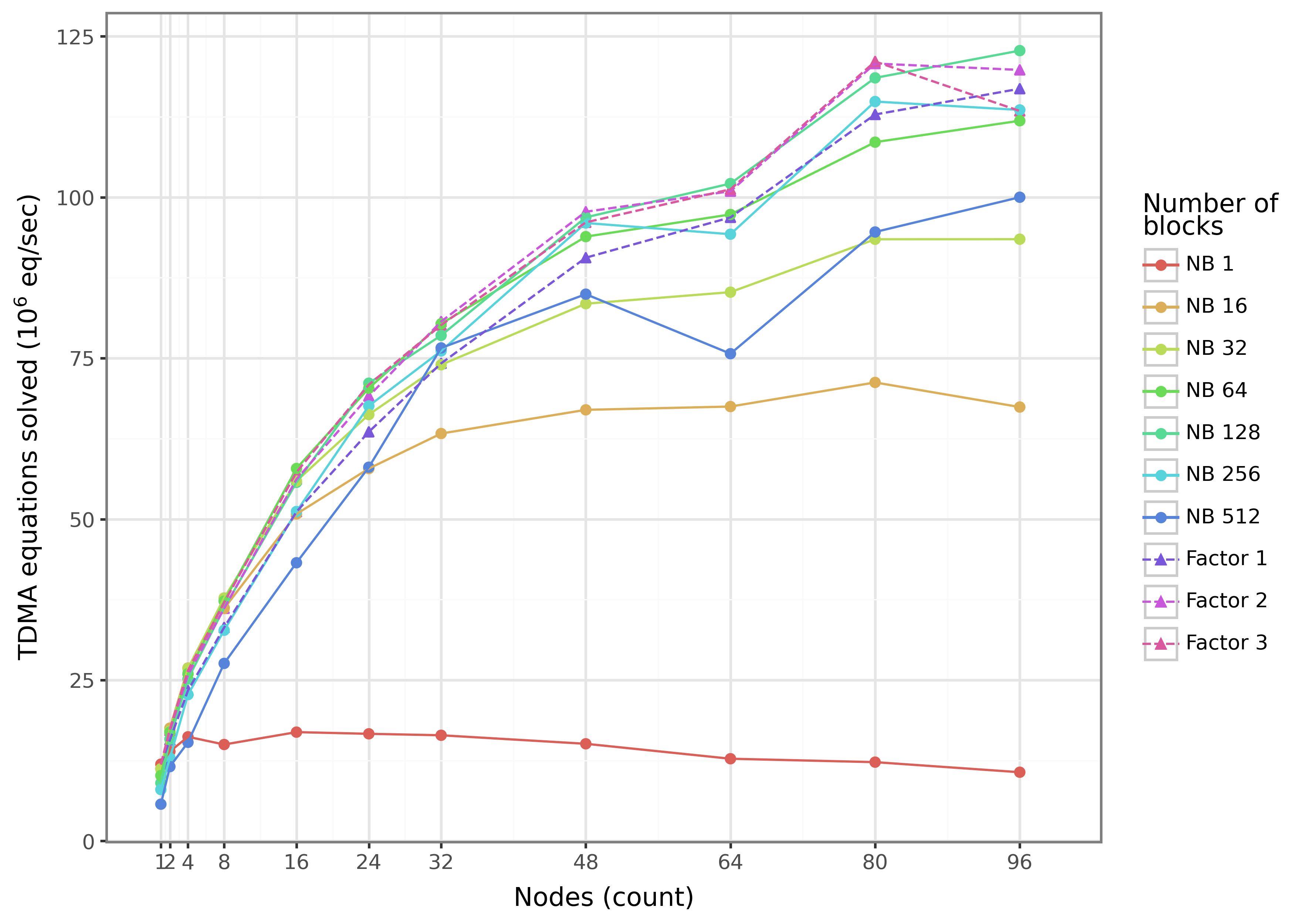}
  \caption{Weak scaling of the BioFVM-B diffusion-decay AVX256D solver. Tests were conducted with increasing workloads and computing nodes, varying the number of blocks. Performance is measured by the number of tridiagonal equation systems processed per second. NB indicates a fixed number of blocks, while Factor represents the ratio between the number of blocks and nodes.} 
  \label{fig:weakscaling}
\end{figure}

We wanted to study the effect of NB in a weak-scaling scenario. Thus, we tested the BioFVM-B diffusion-decay solver with an increasing number of nodes while maintaining the same proportion of work per node and varying NB. The purpose was to observe a sustained increase in computation speed as the number of computing resources and NB increased. Figure \ref{fig:weakscaling} uses as a metric the number of TDMA equations processed per second as the number of computing nodes increases. We observed that low NB values (1, 2, and 4) do not improve computation speed. The highest speeds were obtained with NB set to 128 with factor 1 and 2, solving more than 120 million TDMA equations with 96 computing nodes on the MareNostrum 5 supercomputer. Similar speeds were also reached with NB set to 256 and 64. Lower NB values (1–32) do not exploit parallelism as effectively as the higher ones.\\

Increasing NB enhances the potential parallelism of the diffusion-decay solver. A solver using a single block suffers from serialization, mimicking the behavior of BioFVM-X, and achieves only about 20\% parallel efficiency. As the number of blocks increases, execution time decreases, yielding parallel efficiency of approximately 66\% with 32 and 64 blocks. Figure \ref{fig:blocking} shows execution traces with increasing numbers of blocks and illustrates BioFVM-B’s enhanced parallelization through concurrent communication management and faster workload propagation across all computing nodes.\\

\begin{figure}[h]
  \centering
  \includegraphics[width=\columnwidth]{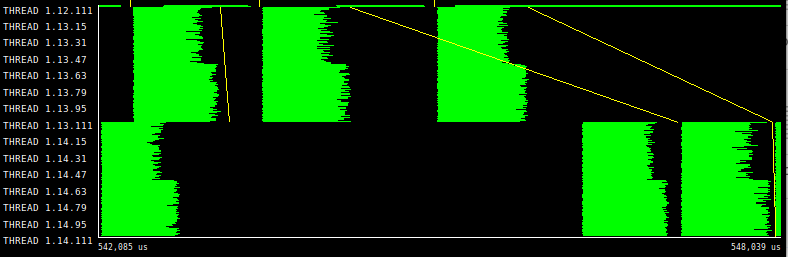}
  \caption{BioFVM-B diffusion-decay AVX256D preemption effect on the block processing. Green regions illustrate parallel regions and yellow lines communication patterns. Zoomed-in showing 13th and 14th nodes of a 32 nodes execution.}
  \label{fig:preemption}
\end{figure}

Despite this, operating system preemption has been observed as a source of uncertainty when determining the most efficient number of blocks for a task. Figure \ref{fig:preemption} shows an execution trace using a large number of nodes, where inefficiencies caused by processes being preempted by the operating system result in delays in block propagation. Although this does not degrade overall performance, it becomes more relevant when large numbers of nodes are combined with a high number of blocks. Due to runtime variability, the optimal number of blocks cannot be determined with certainty. However, we have observed consistent performance using twice as many blocks as nodes (a factor 2 in Equation \ref{eq:blocks}), and this approach is used in the following multi-node scalability tests.\\

\subsubsection{Multi-node performance and scalability}

Multi-node scalability of the diffusion-decay solver was evaluated across various problem sizes with an increasing number of nodes. Each MPI process utilized all 112 cores of a single computational node, following the optimal configuration identified earlier.\\

Columns 6–8 of Table \ref{tab:results} present the execution times per single time step for representative supported problem sizes. BioFVM-B demonstrates good scalability, maintaining execution times under one second for problem sizes up to 21\% of a typical liver, achieving approximately 830 ms per iteration. Simulating the full liver (100\%) results in execution times exceeding 32 seconds per iteration. Although this may seem considerable for a single iteration, it is the first time a full organ environment has been simulated, potentially opening new avenues for disease treatment. \\

In contrast, the BioFVM-X diffusion-decay solver experiences slowdowns as both the problem size and the number of required nodes increase. For the 4\% liver problem, BioFVM-B achieves speedups of 2.24× and 4.36× compared to BioFVM and BioFVM-X, respectively. As the domain size grows, the performance gap between BioFVM-X and BioFVM-B widens, with BioFVM-B being approximately 196.8× faster in the largest test case, 21\% liver size in Table \ref{tab:results}, which represents the maximum size manageable by the other methods.\\

\begin{figure}[h]
  \centering
  \includegraphics[width=\columnwidth]{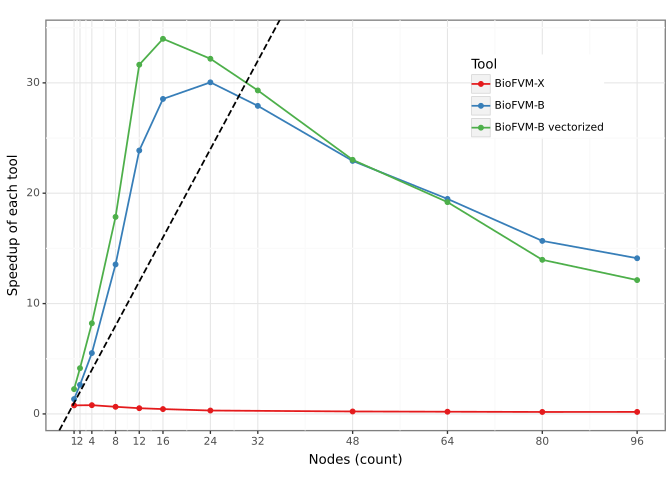}
  \caption{Execution time acceleration of the diffusion-decay solver when modeling a cubic microenvironment equivalent to 4\% of a liver. The reference time corresponds to the BioFVM implementation, and BioFVM-B uses a blocking factor of 2.}
  \label{fig:mn_sc}
\end{figure}

Finally, multi-node strong scaling results demonstrate that the vectorized BioFVM-B is the most performant implementation. Figure \ref{fig:mn_sc} presents the acceleration results, showing that BioFVM-B with vector operations achieves nearly a 35× speedup using 24 computing nodes. Beyond this point, the problem size constrains strong scalability.\\

\subsection{End-to-End Simulation Benchmark}
Besides the isolated solver improvements presented in this work, we analyzed a typical run of BioFVM that includes substrate production, substrate diffusion, and uptake by cells \cite{Ghaffarizadeh2016}. This use case follows six core functions: (1) Resizing of the microenvironment: initializing the \textit{Microenvironment} class with specified parameters; (2) Gaussian profile generation: creating a Gaussian distribution to set initial substrate densities; (3) Initial data storage: saving an initial matrix of substrate values in MATLAB format; (4) Basic agent creation: initiating basic agent creation, assigning them to subdomains, and determining their spatial distribution; (5) Simulation and evolution: solving diffusion-decay equations for substrates as well as uptake and production by cells; and (6) Final data storage: storing the final state of the microenvironment.\\

The main parameters for this experiment include a cubic microenvironment with a side length of 5 mm, containing a single substrate and 1,000 basic agents, 500 functioning as uptake sinks and 500 as bulk sources. The simulation is composed of 500 diffusion time steps, equivalent to 5 minutes of simulated time. The additional BioFVM-B parameters are set to a blocking \textit{factor} of 2 and automatic compiler vectorization.\\

\begin{figure*}[h!]
    \centering
    \includegraphics[width=\textwidth]{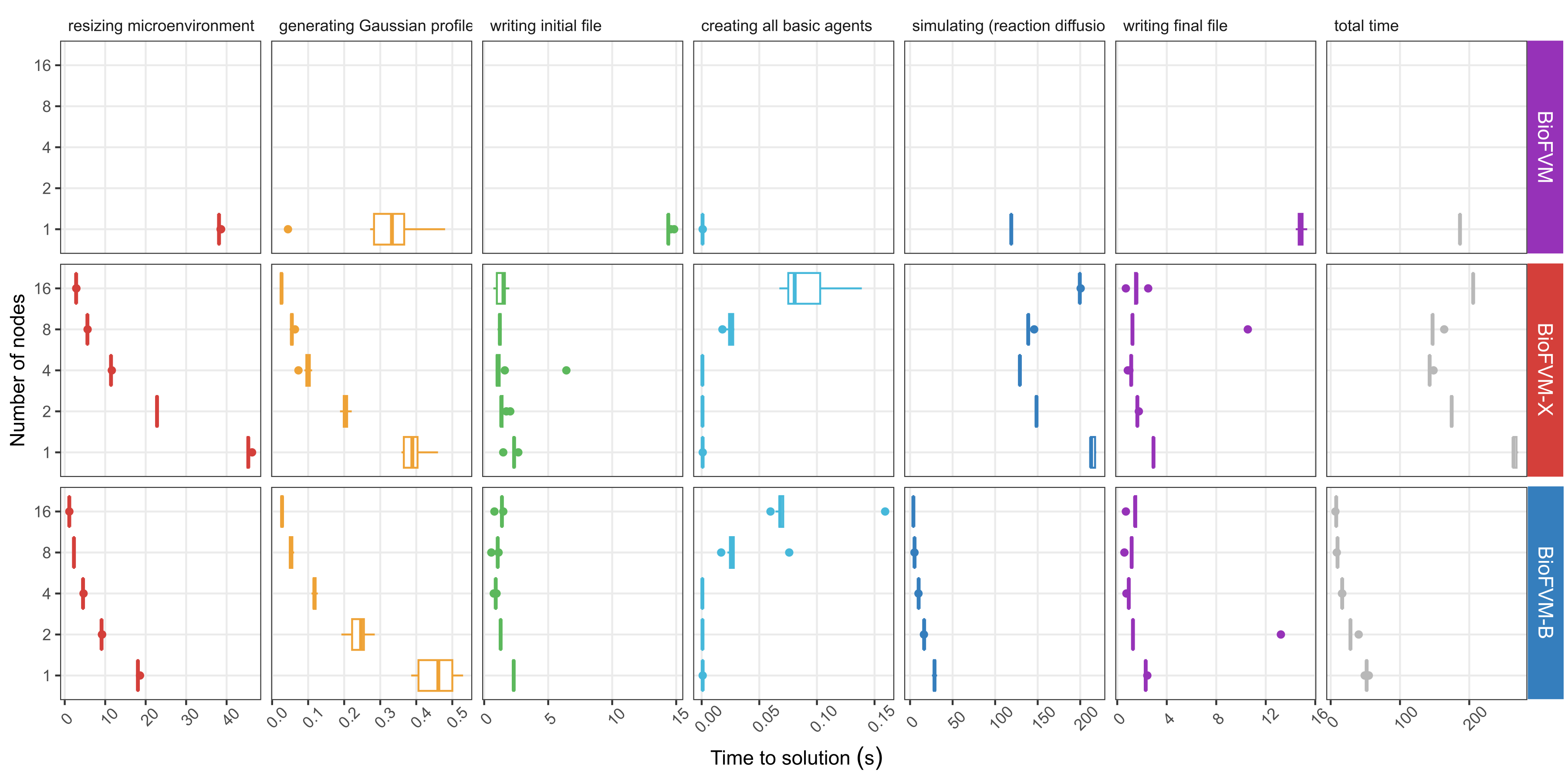}     
    \caption{Ent-to-End simulation benchmark time to solution comparison between BioFVM, BioFVM-X and BioFVM-B.}
    \label{fig:main_experiment}
\end{figure*}

Figure \ref{fig:main_experiment} shows that the stages contributing the most to the total simulation time are, in order: (1) Simulating, (2) Resizing the Microenvironment, and (3) Writing the Initial and Final Concentration Files. The Resize Microenvironment function is primarily influenced by memory allocation and the initialization of auxiliary structures (e.g., Moore neighborhood structures). For this stage, BioFVM-B, featuring optimized methods for this initialization, demonstrates faster single-node execution performance and maintains more efficient scaling with additional nodes than the other libraries. Moreover, the Simulation function, which BioFVM-B is specifically designed to optimize, shows the greatest improvement. As shown in Figure \ref{fig:main_experiment}, BioFVM-X exhibits scalability inefficiencies that are resolved in BioFVM-B, thereby achieving better performance with increasing computational resources. Overall, BioFVM-B outperforms BioFVM and BioFVM-X in both execution time and scalability, proving the beneficial impact of the contributions presented.\\

\section{Conclusions}
The present work introduces an open-source finite volume method (FVM) library optimized for large-scale simulations. BioFVM-B outperforms existing solutions in single and multi-node executions in a supercomputer cluster and is able to scale microenvironment domains, being able to simulate full liver case. Specifically, we solved the performance bottlenecks of the distributed version, BioFVM-X, without altering neither the specification of the solver nor the domain partition enabling large scale simulations that were not feasible previously.\\

BioFVM-B achieves a relevant milestone by enabling the simulation of the diffusion component of a full-organ cellular environment. This advancement marks an important step forward for the field, as it enables agent-based simulators and their extensions to tackle a new class of complex biological problems.\\

Future work will target solver versions for different accelerators, like GPU, FPGA and extended vector machines (notably the RISC-V vector extension) that can better exploit the optimizations presented. In addition and even though the \textit{factor} approach is scalable and reports excellent performance, we will consider elaborating heuristics to find the optimal NB based on the simulation set-up. Finally, BioFVM-B will be used as the cornerstone to enable the distributed computing of agent-based cell simulations using PhysiCell.\\ 

\section*{CRediT authorship contribution statement}
\textbf{José-Luis Estragués-Muñoz:} Conceptualization, Methodology, Software, Validation, Investigation, Writing - Original Draft, Writing - Review \& Editing, Visualization.
\textbf{Carlos Alvarez:} Conceptualization, Methodology, Writing - Original Draft, Writing - Review \& Editing, Funding acquisition. 
\textbf{Arnau Montagud:} Conceptualization, Methodology, Writing - Original Draft, Writing - Review \& Editing, Visualization, Project administration, Funding acquisition.
\textbf{Daniel Jimenez-Gonzalez:} Conceptualization, Methodology, Writing - Original Draft, Writing - Review \& Editing.
\textbf{Alfonso Valencia:} Conceptualization, Writing - Original Draft, Writing - Review \& Editing, Supervision, Funding acquisition.

\section*{Declaration of competing interest}

The authors declare that they have no known competing financial interests or personal relationships that could have appeared to influence the work reported in this paper.

\section*{Acknowledgements}
This work has received funding from the Huawei-BSC Research Agreement, Generalitat de Catalunya (2021 SGR 01007), the Horizon 2020 project PerMedCoE (ID: 951773), the Digital Europe programme project EDITH-CSA (ID: 101083771) and the Horizon Europe projects: CREXDATA (ID: 101092749) and Hanami (ID: 101136269).
AM acknowledges funding from the Generalitat Valenciana’s CIDE GenT programme under the project CIDEXG/2023/22.

\section*{Data availability}
Code and research data are available in the repository 
\hyperlink{https://github.com/bsc-life/BioFVM-B.git}{https://github.com/bsc-life/BioFVM-B.git}

\appendix
\section{BioFVM TDMA's precomputed coefficients} \label{cap:a_coefficients}
Suppose systems of equations of the form \textit{$a_i + b_ix_i + c_i = d_i$} where \textit{n} the number of voxels in a specific dimension, when applied the diffusion-decay solver:
\begin{center}
\begin{equation}
a_{i} =  \frac{dt*\Vec{D}}{dx^2} \; \;  1 \leq i \leq n
\end{equation}
\end{center}

\begin{center}
\begin{equation}
b_{i} = 
\begin{cases} 
1 + \frac{dt\Vec{D}}{dx^2} + \frac{dt\Vec{\lambda}}{3} & i=1 \: or \: i=n\\
1 + \frac{2dt\Vec{D}}{dx^2} + \frac{dt\Vec{\lambda}}{3} & 2 \leq i \leq n-1 \\
\end{cases}
\end{equation}
\end{center}

\begin{center}
\begin{equation}
c_{i} = -\frac{dt*\Vec{D}}{dx^2} \; \;  1 \leq i \leq n
\end{equation}
\end{center}

Coefficients can be pre-computed since Diffusion ($\Vec{D}$) and Decay ($\Vec{\lambda}$) coefficients remain constant during the entire simulation. The first solver iteration performs the simplification and stores the constant coefficients for future iterations, reducing the required number of operations. The precomputed constants are \textit{$constant\_c$} from equation \ref{eq:cprime} and \textit{denom} from equation \ref{eq:denom}:

\begin{equation} \label{eq:cprime}
constant\_c_{i} = 
\begin{cases} 
\frac{c_{i}}{b_{i}} &  i=1\\
\frac{c_{i}}{b_{i} - a_{i}c^{\prime}_{i-1}} & 2 \leq i \leq n \\
\end{cases}
\end{equation}

\begin{center} 
\begin{equation} \label{eq:denom}
denom_{i} =
\begin{cases}
b_{i} & i = 1 \\
b_{i}-a_{i}c^{\prime}_{i-1} & 2 \leq i \leq n
\end{cases}
\end{equation}
\end{center}

\section{BioFVM-B vs. BioFVM-X memory reduction} \label{cap:a_reduction}
\begin{figure}[H]
  \centering
  \includegraphics[width=\columnwidth]{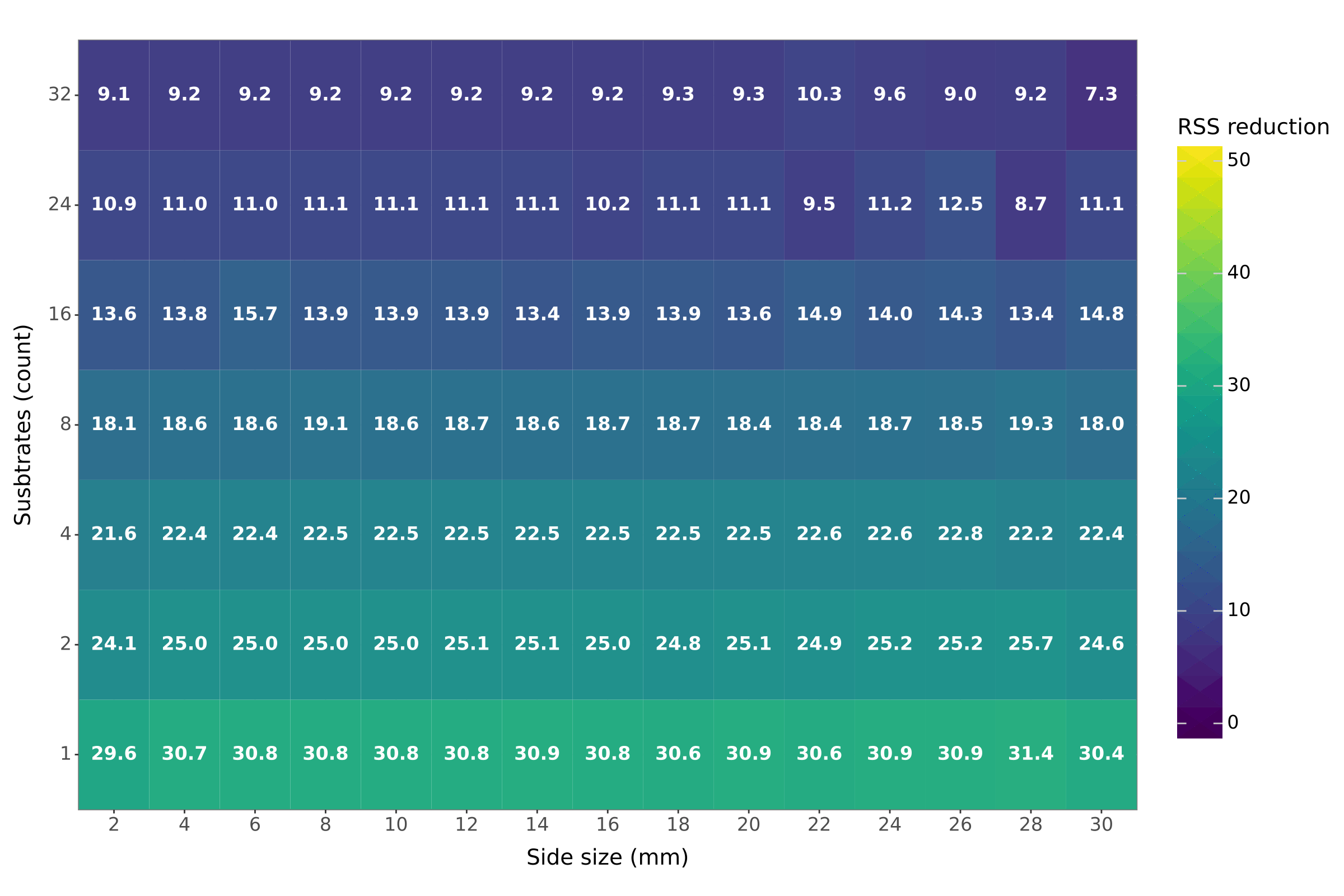}
  \caption{Memory usage reduction in percentage (\%) of using BioFVM-B instead of BioFVM-X.}
  \label{fig:mem_reduction}
\end{figure}

The reduction in memory usage has been observed to correlate with the number of substrates as shown in Figure \ref{fig:mem_reduction}, ranging from approximately 15\% when modeling 32 substrates to around 40\% when modeling only a single substrate, compared to BioFVM-X. \\

\section{BioFVM-B diffusion-decay solver communication and computation overlap} \label{cap:a_blocks}

\begin{figure}[H]
  \centering
  \includegraphics[width=\columnwidth]{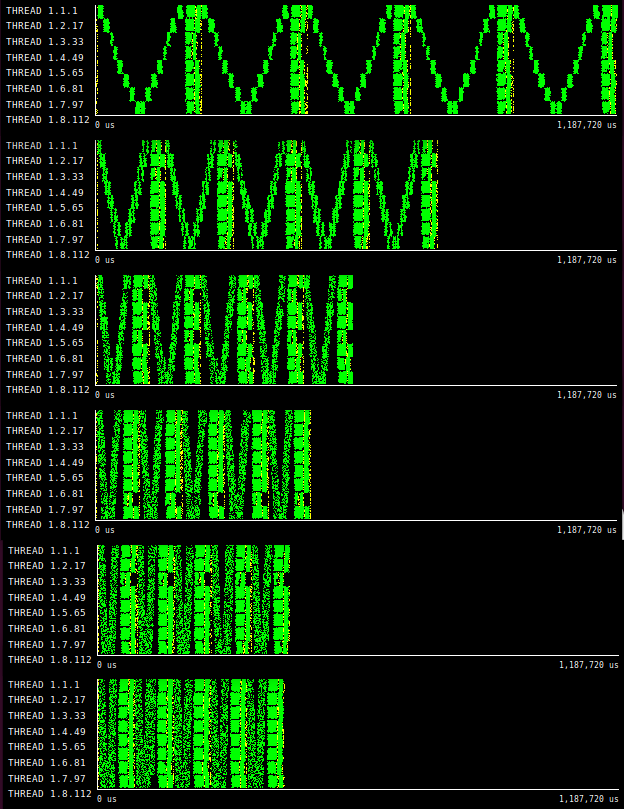}
  \caption{BioFVM-B diffusion-decay AVX256D solver execution traces with increasing number of blocks. Executions using 8 nodes from MareNostrum 5 supercomputer and plotting parallel regions events from 5 executions of the solver. From top to bottom: 1 block, 2 blocks, 4 blocks, 8 blocks, 16 blocks and 32 blocks.}
  \label{fig:blocking}
\end{figure}











\end{document}